\documentclass[twocolumn,amsmath,amssymb,aps,prl,superscriptaddress]{revtex4}
\usepackage{graphicx}
\usepackage{amsmath}
\usepackage{color}
\usepackage{array}
\usepackage{subeqnarray}
\usepackage{ulem}
\usepackage[english]{babel}
\usepackage{setspace}
\usepackage{hyperref}
\usepackage{cleveref}
\usepackage{bm}
\begin{document}
\title{Contactless rheology of finite-size air-water interfaces}
\author{Vincent Bertin}
\thanks{These authors contributed equally.}
\affiliation{Univ. Bordeaux, CNRS, LOMA, UMR 5798, 33405 Talence, France.}
\affiliation{UMR CNRS Gulliver 7083, ESPCI Paris, PSL Research University, 75005 Paris, France.}
\author{Zaicheng Zhang}
\thanks{These authors contributed equally.}
\affiliation{Univ. Bordeaux, CNRS, LOMA, UMR 5798, 33405 Talence, France.}
\author{Rodolphe Boisgard}
\affiliation{Univ. Bordeaux, CNRS, LOMA, UMR 5798, 33405 Talence, France.}
\author{Christine Grauby-Heywang}
\affiliation{Univ. Bordeaux, CNRS, LOMA, UMR 5798, 33405 Talence, France.}
\author{Elie Rapha\"{e}l}
\affiliation{UMR CNRS Gulliver 7083, ESPCI Paris, PSL Research University, 75005 Paris, France.}
\author{Thomas Salez}
\email{thomas.salez@u-bordeaux.fr}
\affiliation{Univ. Bordeaux, CNRS, LOMA, UMR 5798, 33405 Talence, France.}
\affiliation{Global Station for Soft Matter, Global Institution for Collaborative Research and Education, Hokkaido University, Sapporo, Hokkaido 060-0808, Japan.}
\author{Abdelhamid Maali}
\email{abdelhamid.maali@u-bordeaux.fr}
\affiliation{Univ. Bordeaux, CNRS, LOMA, UMR 5798, 33405 Talence, France.}
\date{\today}

\begin{abstract}
We present contactless atomic-force microscopy measurements of the hydrodynamic interactions between a rigid sphere and an air bubble in water at the micro-scale. The size of the bubble is found to have a significant effect on the response due to the long-range capillary deformation of the air-water interface. To rationalize the experimental data, we develop a viscocapillary lubrication model accounting for the finite-size effect. The comparison between experiments and theory allows us to measure the air-water surface tension, without contact, paving the way towards robust contactless tensiometry of polluted air-water interfaces.
\end{abstract}

\maketitle

The interface between two media has an energy cost per unit surface, called surface tension, resulting from the microscopic interactions of the constitutive molecules at the interface~\cite{de2013capillarity,marchand2011surface}. Surface tension is an important parameter in soft condensed matter and at small scales where capillary phenomena usually dominate. Examples include wetting properties~\cite{de1985wetting,bonn2009wetting}, thin-film dynamics~\cite{oron1997long,craster2009dynamics}, multiphase flows...

Surface active molecules -- \textit{i.e.} surfactants -- are widely used to stabilize capillary interfaces on purpose, \textit{e.g.} in emulsions or foams, but are also inevitable due to pollution. These contaminants, which are usually adsorbed at the interface between two immiscible fluids, lower the surface tension and are responsible for specific rheological properties of the interface~\cite{langevin2014rheology}. To understand the dynamics of soft materials, the interaction between objects such as droplets and bubbles, or to quantify the amount of interfacial contamination, capillary interfacial rheology is essential. Specifically, surface tension is measured by a large variety of techniques: pendant-drop method~\cite{berry2015measurement}, spinning-drop method, Wilhelmy plates or du No\"uy rings~\cite{drelich2002measurement}, for instance. Moreover, the interfacial rheology is usually measured with the Langmuir trough~\cite{schwartz1994direct} or through oscillating-disk devices~\cite{erni2003stress}.

Complementary devices to measure material properties are atomic force microscope (AFM) and surface forces apparatus (SFA), which have recently been used to study capillary phenomena such as the interaction between bubbles~\cite{vakarelski2008bubble,vakarelski2010dynamic} or droplets~\cite{dagastine2006dynamic,chan2011film,tabor2012measurement}, the hydrodynamic boundary condition at a water-air interface~\cite{steinberger2007high,manor2008dynamic,maali2017viscoelastic,wang2018viscocapillary}, and dynamical wetting~\cite{ecke1999microsphere,xiong2009development,delmas2011contact,guo2013direct,dupre2015shape,mortagne2017dynamics}. Recently, AFM and SFA have also been employed in dynamical modes, and appear to be remarkable tools in order to quantify material properties -- with the advantage of providing contactless measurements~\cite{leroy2011hydrodynamic,leroy2012hydrodynamic,villey2013effect,guan2017noncontact,Zhang2020}.

In this Letter, we study the force exerted on a water-immersed sphere attached to an AFM cantilever, that is driven to oscillate near the apex of an air bubble. The deformation of the bubble and the force exerted on the spherical probe are coupled, and result from the hydrodynamic pressure induced by the oscillating water flow. To rationalize the experimental data, we develop a lubrication model accounting for finite-size effects -- which are found to be significant in the linear viscocapillary response. All together, this method allows for robust interfacial rheology in the absence of any direct contact. 
\begin{figure}[t!]
\centering
\includegraphics[width=0.8\columnwidth]{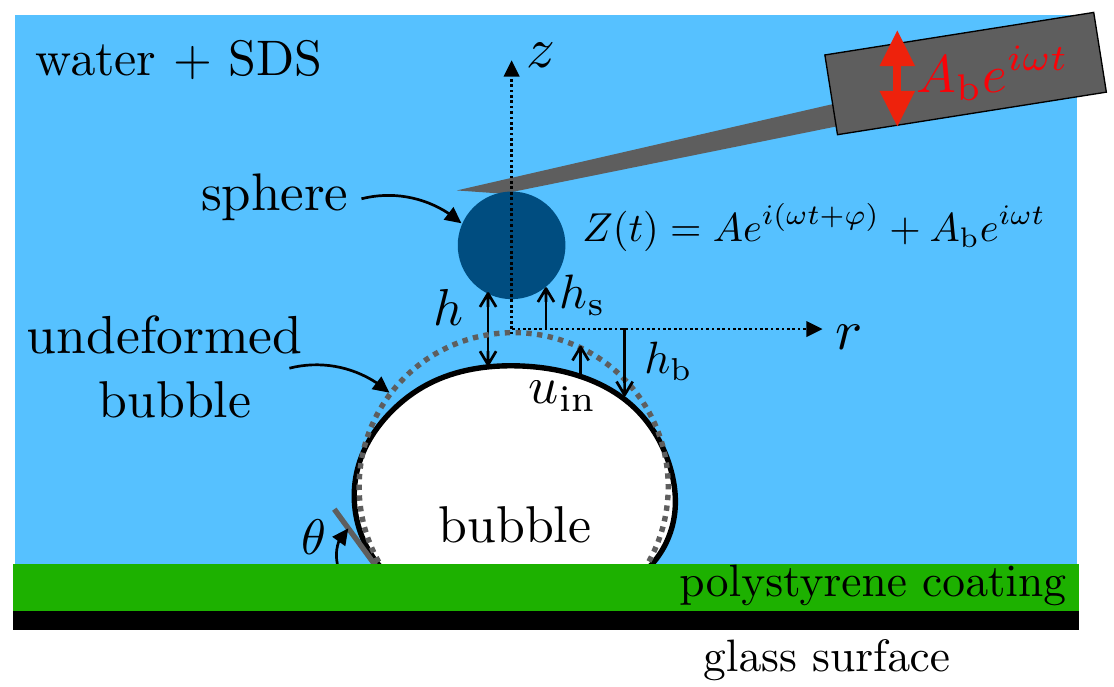}
\caption{A spherical probe (dark blue) attached to an AFM cantilever (grey) oscillates vertically, along the $z$ axis, within a water-Sodium Dodecyl Sulfate (SDS) liquid solution (light blue) and near an air bubble (white). The motion results in an axisymmetric liquid-gap thickness profile $h(r,t)$ depending on the radial distance $r$ and time $t$, that includes an axisymmetric vertical deformation field $u_{\textrm{in}}(r,t)$ of the bubble surface (defined along the $+z$ direction), with respect to its equilibrium spherical-cap shape (dashed line).}
\label{fig:schema}
\end{figure}

The schematic of the experimental setup is shown in Fig.~\ref{fig:schema}~\cite{maali2017viscoelastic, maali2013precise}. The cantilever is excited by the base oscillation $\mathcal{R}[A_\text{b}\text{e}^{i\omega t}]$, where $\omega$ and $A_\text{b}$ are the angular frequency and amplitude of the base vibration, respectively, and $\mathcal{R}[.]$ denotes the real part. The system essentially behaves as a damped oscillator, where the vertical displacement $Z(t)$ of the center of mass of the sphere with respect to its rest position satisfies:
\begin{equation}
m_\text{c} \ddot{Z}+\Gamma_\text{bulk}\dot{Z}+k_\text{c}Z = F_\text{d}+F\ ,
\label{eq:sphere motion}
\end{equation}
with $m_\text{c}$ the effective mass (\textit{i.e.} including the added fluid mass), $\Gamma_\text{bulk}$ the damping coefficient in the bulk, $k_\text{c}$ the stiffness of the cantilever, $F_\text{d}$ the driving force due to the imposed oscillation of the cantilever, and $F = \mathcal{R}[F^* \text{e}^{i\omega t}]$ the hydrodynamic force resulting from the interaction between the oscillating sphere and the air-water interface. 
The displacement  $Z(t)$ of the sphere includes the cantilever deflection $\mathcal{R}[A\text{e}^{i(\omega t +\varphi)}]$ measured by AFM and the base displacement, and thus reads $ Z(t)=\mathcal{R}[A\text{e}^{i(\omega t +\varphi)}+A_\text{b}\text{e}^{i\omega t}] = \mathcal{R}[Z^\ast \text{e}^{i\omega t}]$, where $A$ and $Z^\ast=A\text{e}^{i\varphi}+A_\text{b}$ are real and complex amplitudes respectively~\cite{maali2013precise}. We further define the mechanical impedance $G^\ast=-F^*/Z^*$. Invoking the complex version of Eq.~\eqref{eq:sphere motion}, the impedance reads: 
\begin{equation}
G^\ast =- k_\text{c}\left[1-\left(\frac{\omega}{\omega_0}\right)^2+i\frac{\omega}{\omega_0 Q}\right]\frac{A\text{e}^{i\varphi}-A_\infty \text{e}^{i\varphi_\infty}}{A\text{e}^{i\varphi}+A_\text{b}}\ ,
\label{eq:measured hydrodynamic force}
\end{equation}
where $A_\infty$ and $\varphi_{\infty}$ are respectively the amplitude ($A$) and phase ($\varphi$) measured far from the bubble (\textit{i.e.} where $F$ vanishes), $\omega_0=\sqrt{k_\text{c}/m_\text{c}}$ is the bulk resonance frequency, and $Q=m_\text{c}\omega_0/\Gamma_\text{bulk}$ is the bulk quality factor. Equation~\eqref{eq:measured hydrodynamic force} provides a direct way to measure $G^\ast$ experimentally from the cantilever's deflection signal.

To model theoretically $G^\ast$, we consider the axisymmetric system composed of the rigid sphere located at an average distance $D$ from the apex of the undeformed air bubble. The ensemble is immersed in an incompressible Newtonian fluid with a dynamical shear viscosity $\eta$. We define the effective curvature radius $R_\text{eff}$ of the lubricated contact from: $R_\text{eff}^{-1} = R_\textrm{s}^{-1} + R_\textrm{b}^{-1}$, where $R_\textrm{s}$ and $R_\textrm{b}$ are the curvature radii of the sphere and bubble, respectively. We focus on the situation where $D\ll R_\textrm{eff}$, so that we can invoke the lubrication approximation of the steady Stokes equations. The experiments are done at low enough frequencies so that we can assume a no-slip boundary condition at the air-water interface~\cite{maali2017viscoelastic}. Such a condition is also assumed at the sphere-liquid interface. Therefore, the liquid-gap thickness obeys the Reynolds equation~\cite{Reynolds1886}: 
\begin{equation}
\frac{\partial h(r,t) }{\partial t} = \frac{1}{12\eta r} \frac{\partial}{\partial r}\bigg[rh(r,t)^3\frac{\partial}{\partial r} p(r,t) \bigg]\ ,
\label{eq:lubrication}
\end{equation}
where $p(r,t)$ is the excess hydrodynamic pressure field with respect to the rest state, $h(r,t) = h_\textrm{s}(r,t) - h_\textrm{b}(r,t)$ is the liquid-gap thickness, and $h_\textrm{s}$ and $h_\textrm{b}$ are the sphere and bubble surface profiles respectively (see Fig.~\ref{fig:schema}). The later follows the Young-Laplace equation:
\begin{equation}
\label{eq:Young-Laplace}
\frac{\gamma}{r}\frac{\partial}{\partial r} \left[r \, \frac{\frac{\partial h_\textrm{b}}{\partial r}}{\sqrt{1 + \left(\frac{\partial h_\textrm{b}}{\partial r}\right)^2}} \right] = \Delta P(t) + p(r,t)\ ,
\end{equation}
where $\gamma$ denotes the air-water surface tension, and $\Delta P$ is the pressure drop across the interface. The contribution of Hamaker forces is neglected in the model, as the sphere-bubble distance in the experiment is in the $10$~nm - 20~$\mu$m range, and thus typically larger than the distance below which these forces are dominant. 

Even though the excess hydrodynamic pressure field is essentially localized near the apex of the bubble, over a radial extent that scales with the hydrodynamic radius $\sim \sqrt{2R_\textrm{eff}D}$, the Young-Laplace equation induces deformations on larger scales -- typically the millimetric capillary length~\cite{de2013capillarity}. Here, we consider bubbles with radii $R_\textrm{b}$ smaller than the capillary length, so that: i) gravity is neglected; and ii) the radial extent of the bubble's deformation induced by the hydrodynamic pressure is rather set by $R_\textrm{b}$. Thus, we expect finite-size effects. Interestingly, there is still a natural scale separation since $\sqrt{2R_\textrm{eff}D} \ll R_\textrm{b}$ in practice. We thus use an asymptotic-matching method to solve the problem~\citep{chan2011film}. The outer solution (denoted with subscript ``out") is solved exactly and depends directly on the hydrodynamic force $F(t)=2\pi\int_0^{\infty}\textrm{d}r\,r\,p(r,t)$~\citep{appendix}. We treat the inner hydrodynamic region (denoted with subscript ``in") as a boundary layer of the full interfacial deformation. We further assume a small amplitude of the sphere's oscillation, and we expand the inner surface profile of the bubble as: $h_\textrm{b,in}(r,t) = -r^2/(2R_\textrm{b})-u_\textrm{in}(r,t)$, where $u_\textrm{in}$ is the inner flow-induced capillary deformation of the air-water interface. The linearization of Eq.~\eqref{eq:Young-Laplace} near the apex of the bubble, combined with the air-volume conservation, yields~\citep{appendix}:
\begin{equation} 
\label{eq:Young-Laplace_inner_deformation}
-\frac{\gamma}{r}\frac{\partial}{\partial r} \bigg(r \, \frac{\partial u_{\textrm{in}}}{\partial r} \bigg) = -\frac{F(t)}{\pi R_\textrm{b}^2}\frac{1}{1+\cos\theta} + p(r,t)\ ,
\end{equation}
where $\theta$ is the contact angle (see Fig.~\ref{fig:schema}), and where we assumed the contact line to be pinned. The matching of the inner and the outer solutions imposes the asymptotic behavior $u_\textrm{in}(r,t) \sim -\frac{F(t)}{2\pi \gamma} \left[1 - \log \left(\frac{1+\cos\theta}{1-\cos\theta}\right) + \log\left(\frac{r}{2R_\textrm{b}}\right)  \right]$, for $r\gg \sqrt{2R_\textrm{eff}D}$. We invoke the linear-response framework, and write: $u_\textrm{in}(r,t)=\mathcal{R}[u^\ast_\textrm{in}(r) \text{e}^{i\omega t}]$ and $p(r,t)=\mathcal{R}[p^\ast(r) \text{e}^{i\omega t}]$, with $u^\ast_\textrm{in}(r)$ and $p^\ast(r)$ the corresponding complex amplitudes. Finally, the amplitude of the hydrodynamic force reads $F^*=2\pi \int_0^{\infty}\textrm{d}r\, r\,p^*(r)$ in the lubrication approximation, which allows us to compute the mechanical impedance $G^*$~\citep{appendix}. 

The experiments are performed using an AFM (Bioscope, Bruker) equipped with a liquid cell (DTFML-DD-HE). A spherical borosilicate particle (MO-Sci Corporation) with a $R_\textrm{s}=54 \pm  2~\mu$m radius is glued at the edge of a silicon nitride cantilever (ORC8-10, Bruker AFM Probes). The stiffness $k_\text{c}=0.20 \pm 0.01~\mathrm{N/m}$ of the cantilever (with the sphere attached to it) is determined from the drainage method~\cite{craig2001situ}. The bulk resonance frequency $\omega_0/(2\pi)=1240 \pm  3~\text{Hz}$ and the bulk quality factor $Q=3.4  \pm  0.1$ are obtained from the resonance spectrum at large distance~\cite{maali2013precise}. Air microbubbles are deposited onto spincoated polystyrene layers, within Sodium Dodecyl Sulfate (SDS) solutions in water. The SDS concentrations $C$ are in the $0.2-40\,\textrm{mM}$ range. As measured with an optical microscope, the bubble radii $R_\textrm{b}$ are in the $0.2 - 0.6$~mm range, and the contact angles $\theta$ (see definition in Fig.~\ref{fig:schema}) are in the $40-90^\circ$ range, with the exact value depending on $C$. A multi-axis piezo stage (NanoT series, Mad City Labs) is used to control the distance between the sphere and the bubble, by imposing a displacement to the substrate at very low velocity. The amplitude $A$ and phase $\varphi$ of the cantilever's deflection signal are measured by a lock-in amplifier (Model 7280, Signal Recovery), and are recorded versus the piezo displacement. Additionally, the DC component of the cantilever's deflection is also recorded and used to determine the average gap distance $D$. We stress that the amplitude of the spherical probe's oscillation is a few nanometers, and always less than 3.5~nm, which is itself smaller than $D$, fixed to be in the 10~nm - 20~$\mu$m range.

\begin{figure}[t!]
\centering
\includegraphics[width=\columnwidth]{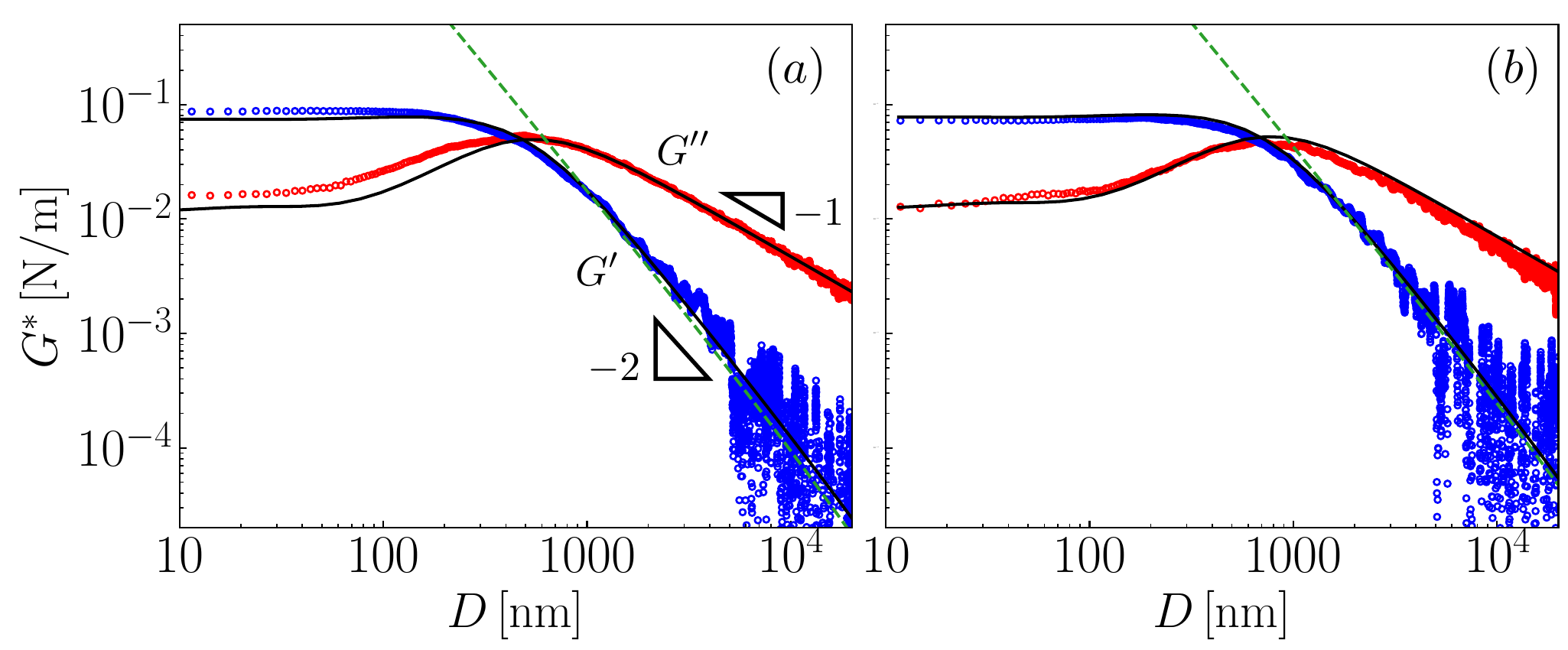}
\caption{Real (blue circles) and imaginary (red circles) parts of the measured mechanical impedance $G^* = G' + i G''$ versus average sphere-bubble distance $D$, for a surfactant concentration $C=1 \, \textrm{mM}$, and frequencies $\omega/(2\pi)=200$~Hz (a), and $300$~Hz (b). The bubble radius is $R_\text{b}=346\pm2~\mu\text{m}$ and the contact angle is $\theta=81\pm2^\circ$. The best fits to the model~\citep{appendix} are displayed with solid black lines, using a single fitting parameter $\gamma = 54\pm4~\textrm{mN}/\textrm{m}$. The large-distance asymptotic solution for $G'$ (see Eq.~\eqref{eq:asymptotic}) is also shown with green dashed lines. The slope triangles indicate power-law exponents.}
\label{fig:impedance}
\end{figure}
The real and imaginary parts of the measured mechanical impedance $G^* = G' + i G''$ are plotted in Fig.~\ref{fig:impedance}, as functions of the average sphere-bubble distance $D$, for two frequencies and a given surfactant concentration. Best fits to the model ~\citep{appendix} are also shown, in good agreement with the data, the air-water surface tension $\gamma = 54\pm4~\textrm{mN}/\textrm{m}$ being the only fitting parameter. Furthermore, two asymptotic regimes can be observed, at large and small distance respectively. They crossover near $D \approx 1000 \, \textrm{nm}$, which corresponds to the typical viscocapillary distance $D_\textrm{c} = 16R_\text{eff}^2 \eta \omega /\gamma$ emerging from the model~\citep{appendix}, and equal to $771$ and $1 160 \, \textrm{nm}$ in Figs.~\ref{fig:impedance}(a) and (b), respectively. At large distance, the viscous contribution $G''$ dominates and follows a $\sim D^{-1}$ scaling law, as expected from the asymptotic expression $G'' \simeq 6\pi \eta R_\textrm{eff}^2\omega/D$~\cite{leroy2011hydrodynamic}. In contrast, the restoring contribution $G'$ due to the air-water capillary interface appears with an apparent $\sim D^{-2}$ scaling law at large distance. We stress that the latter is not an exact scaling law, due to a logarithmic correction~\cite{appendix}: 
\begin{equation}
\label{eq:asymptotic}
\begin{split}
G'(D) \simeq \frac{9\pi \eta^2 R_\textrm{eff}^4\omega^2}{\gamma D^2}\, \bigg[&-3+\log( 4) - 2\log\left( \frac{1+\cos\theta}{1-\cos\theta}\right) \\
&+ \log\left(\frac{R_\textrm{b}^2}{2R_\textrm{eff}D}\right) \bigg]\ .
\end{split}
\end{equation}
At small distance, both $G'$ and $G''$ saturate to constant values, which is reminiscent of elastohydrodynamic responses near soft substrates~\cite{Skotheim2005,leroy2011hydrodynamic, leroy2012hydrodynamic, villey2013effect,Saintyves2016,Zhang2020}, and might be related to saturations in the deformation and pressure fields. At such small distances, the capillary deformation of the bubble surface essentially accommodates the sphere's oscillation, and the liquid is no longer expelled from the gap, which further leads to a stronger capillary response than the viscous one.

\begin{figure}[t!]
\centering
\includegraphics[width=\columnwidth]{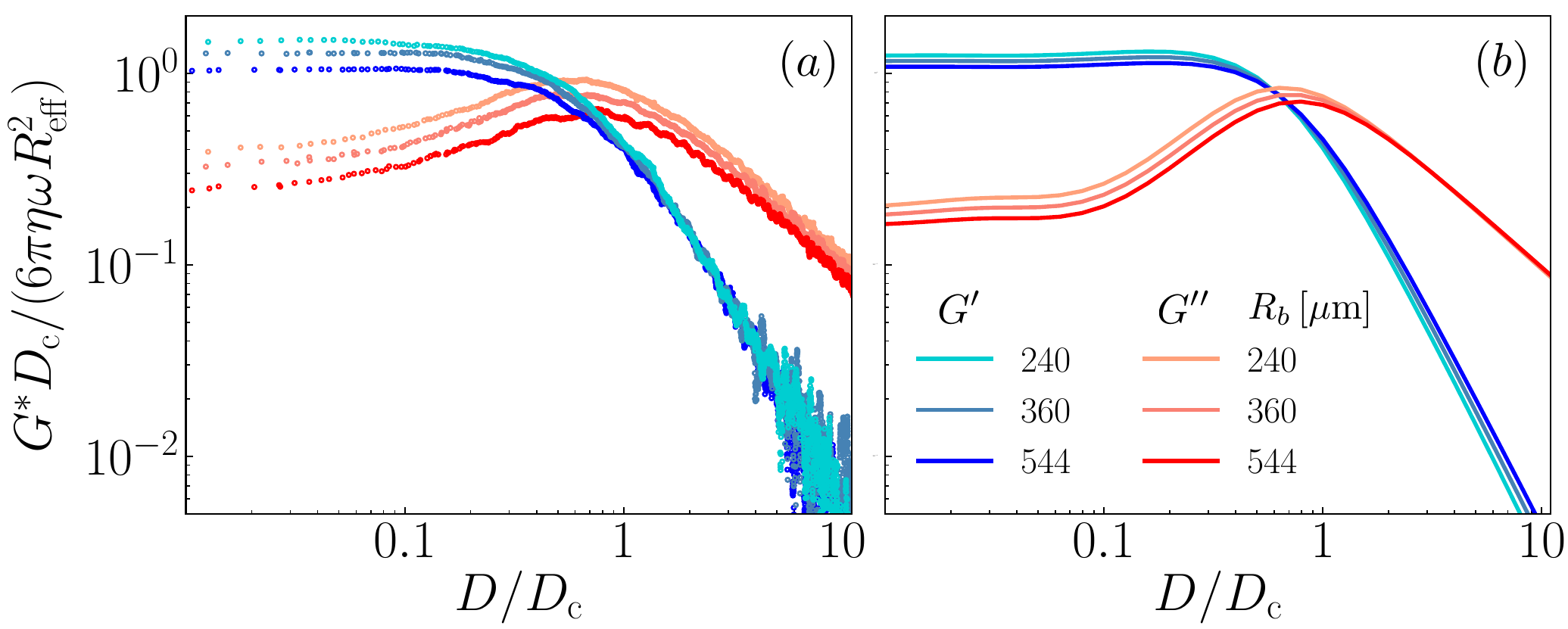}
\caption{Dimensionless mechanical impedance versus dimensionless distance, for three bubble sizes as indicated, a single frequency $\omega/(2\pi)=200$~Hz, and a single surfactant concentration $C=1\,\textrm{mM}$. The experimental data are show in (a). The results of the model are plotted in (b), using the previously-obtained best-fit parameter $\gamma = 54~\textrm{mN}/\textrm{m}$.}
\label{fig:impedance_bubble_size}
\end{figure}

In order to reveal the importance of finite-size effects in the viscocapillary response, we introduce the dimensionless mechanical impedance $\mathcal{G}^*=G^*D_{\textrm{c}}/(6\pi \eta \omega R_\text{eff}^2)$. In Fig.~\ref{fig:impedance_bubble_size}, the experimental and theoretical dimensionless mechanical impedances are plotted versus the dimensionless average sphere-bubble distance, for three bubble radii. Except for the viscous contribution in the large-distance limit, the dimensionless impedance is generally found to depend on the bubble size in a nontrivial way, which is correctly reproduced by the model. This observation highlights the importance of finite-size effects in viscocapillary interactions, resulting from the long-range capillary deformation of the air-water interface. We note that the logarithmic correction in the large-distance asymptotic expression of the capillary contribution (see Eq.~\eqref{eq:asymptotic}) contains a bubble-size dependence which cannot be resolved with the AFM sensitivity and the current bubble-size range. At small distance, the size dependence is more pronounced and both the real and imaginary parts of the dimensionless impedance decrease when increasing the bubble size. 

\begin{figure}[t!]
\centering
\includegraphics[width=\columnwidth]{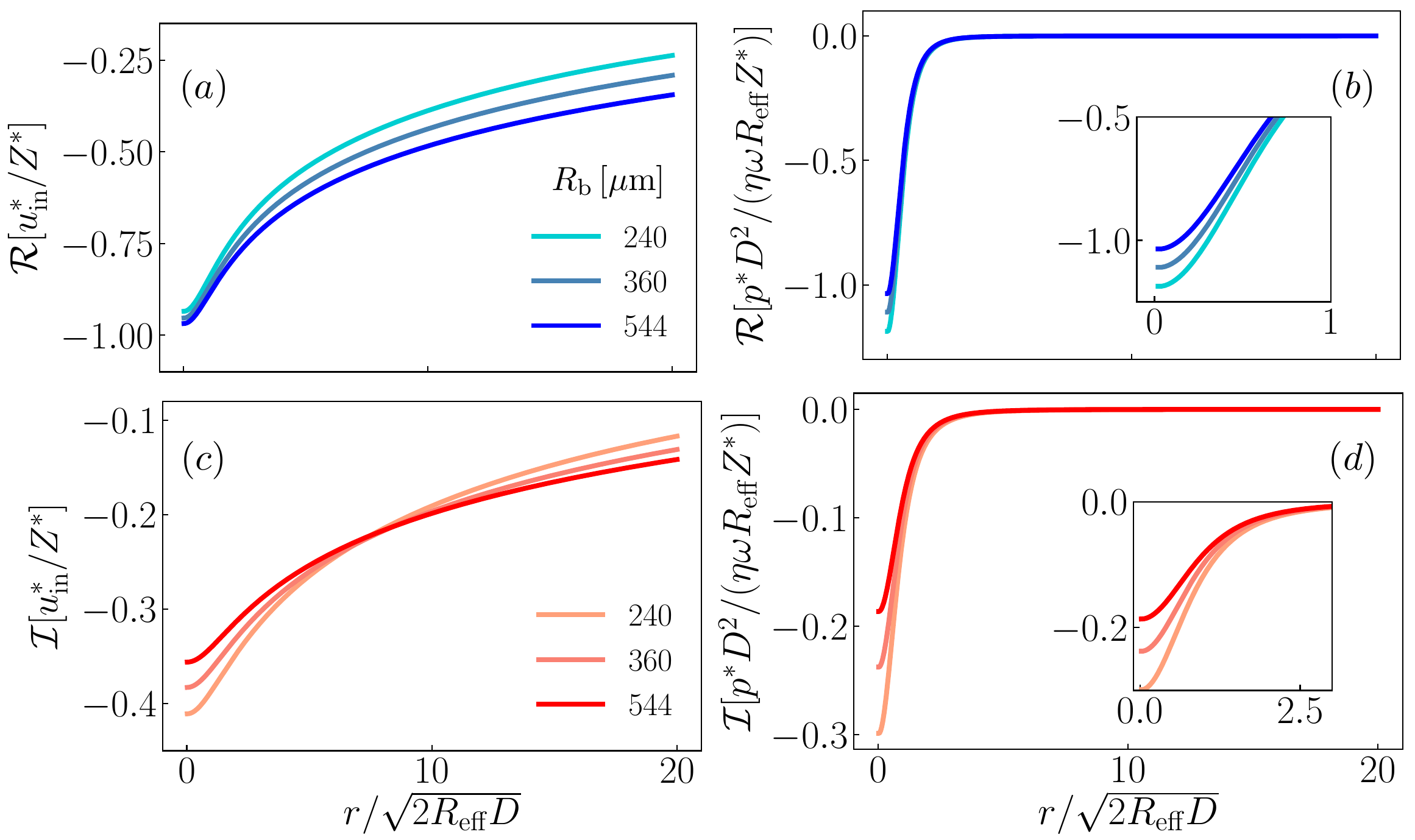}
\caption{Real (a) and imaginary (c) parts of the amplitude of the dimensionless inner deformation field as functions of the dimensionless radial coordinate, at a dimensionless distance $D/D_\textrm{c} = 0.3$, for the three bubble radii of Fig.~\ref{fig:impedance_bubble_size}, as obtained from the model~\citep{appendix}. Similarly, the real and imaginary parts of the amplitude of the dimensionless excess pressure field are plotted in panels (b) and (d), respectively. The insets display zooms near the symmetry axis.}
\label{fig:pressure_deplacement_fields}
\end{figure}

Having discussed the finite-size effects on the global hydrodynamic force, we now investigate their influence on the amplitudes of the local excess pressure and deformation fields. To do so, we perform numerical integrations of Eqs.~\eqref{eq:lubrication} and~\eqref{eq:Young-Laplace_inner_deformation} using the asymptotic expression for the inner deformation field~\cite{appendix}. Figure~\ref{fig:pressure_deplacement_fields} shows the results for $D/D_\textrm{c} = 0.3$, with the same parameters as in Fig.~\ref{fig:impedance_bubble_size}. We observe that the real and imaginary parts of the dimensionless amplitude of the excess pressure field decay rapidly on a typical distance $\sim\sqrt{R_\textrm{eff}D}$, and depend weekly on the bubble radius. In sharp contrast, the real and imaginary parts of the dimensionless amplitude of the inner deformation field largely depend on the bubble radius, as well as on the contact angle (not shown). 

\begin{figure}[t!]
\centering
\includegraphics[width=0.8\columnwidth]{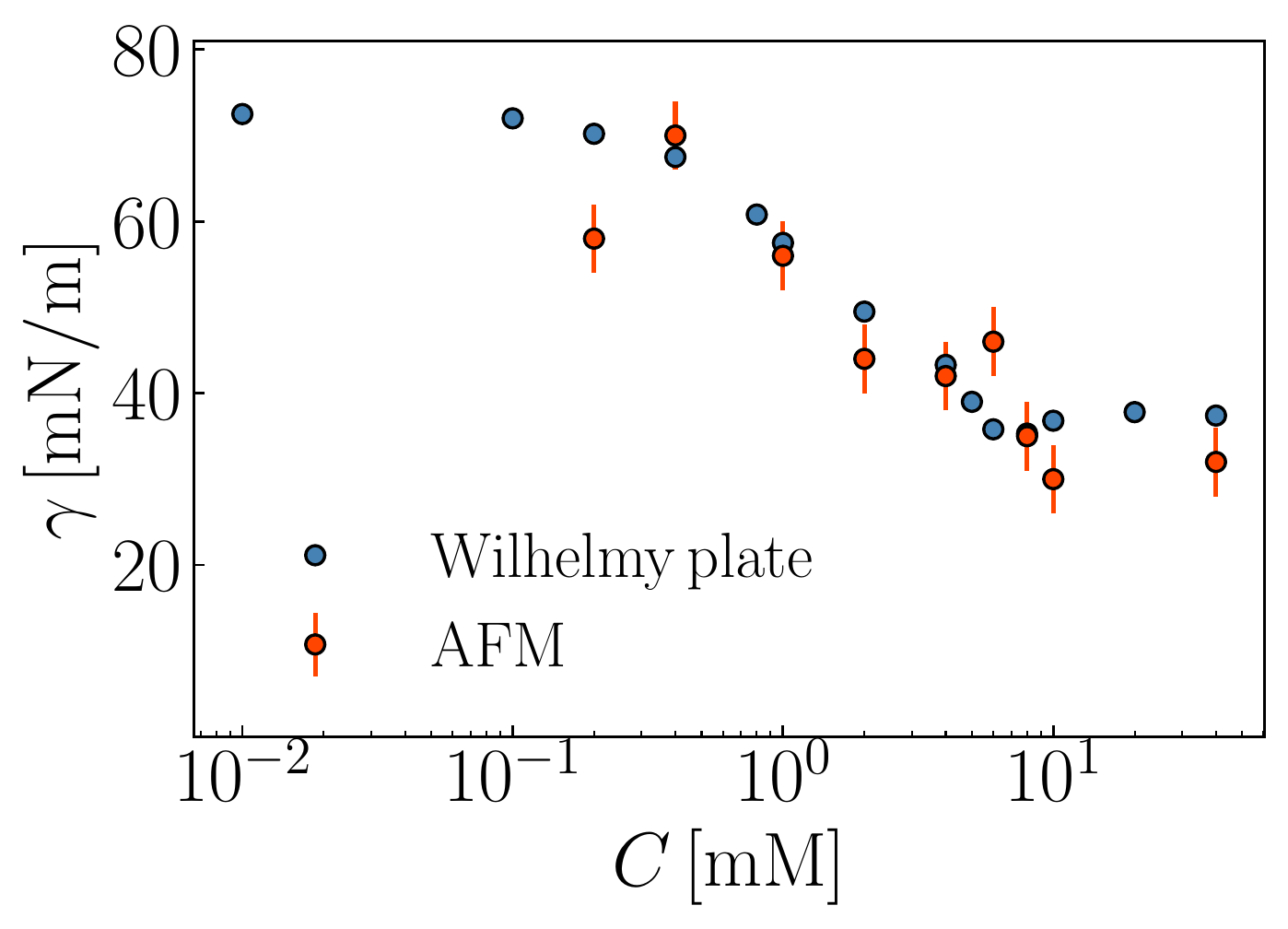}
\caption{Air-water surface tension as a function of surfactant (SDS) concentration, as obtained from fits (see Fig.~\ref{fig:impedance}) of the AFM experimental data by the model (red dots). Statistical error bars associated with multiple measurements at different frequencies are indicated. For comparison, independent measurements using the Wilhelmy-plate method are provided (blue dots).}
\label{fig:surface_tension}
\end{figure}

So far, the air-water surface tension was considered as a free parameter and was fixed by fitting the AFM experimental data to the model. The fitted values of the surface tension as a function of the SDS concentration in water are shown in Fig.~\ref{fig:surface_tension}. We observe that the surface tension globally decreases with increasing surfactant concentration, as expected. At surfactant concentrations smaller than $\sim0.5 \ \textrm{mM}$, the surface tension is close to the $72 \ \textrm{mN/m}$ value for pure water. At concentrations larger than $\sim8\ \textrm{mM}$, the surface tension saturates to a value on the order of $30~\textrm{mN/m}$. The critical micellar concentration of SDS in water is estimated to be around $8\ \textrm{mM}$~\cite{moroi1974critical,fuguet2005critical}, which is in agreement  with the latter observation. The uncertainty on the fitted values of the surface tension is on the order of $\pm 4\ \textrm{mN/m}$, which mainly results from the fact that the experiments at different frequencies lead to slight variations.

Finally, we discuss the capacity of our method to be used as a robust tensiometer. To do so, we perform independent tensiometry experiments on similar air-water-SDS interfaces using the Wilhelmy-plate method~\cite{drelich2002measurement}. The results are shown in Fig.~\ref{fig:surface_tension}, and agree well with the ones obtained with our method. Possible systematic deviations at the highest concentrations may result from a surfactant-induced depinning of the contact line of the bubble on the substrate~\cite{joshi2020effect}. In such a scenario, the hydrodynamic pressure would not only trigger a local capillary deformation (see Eq.~\eqref{eq:Young-Laplace}), but would also induce a spreading-dewetting cycle of the bubble on the substrate. In addition, the bubble resonance frequency being lower at lower surface tension, capillary waves might be excited at the air-water interface at large surfactant concentrations. Besides, at the smallest concentration (0.2 mM) used in the AFM experiment, the air-water interface may not be entirely covered with an adsorbed surfactant layer, potentially resulting in slippage. In such a scenario, the surfactant advection induced by the flow would add an elastic component to the mechanical response~\cite{manor2008dynamic,maali2017viscoelastic}, which might explain the small deviation observed in Fig.~\ref{fig:surface_tension}. 

In conclusion, we have studied the viscocapillary interaction between an air bubble and a spherical probe attached to an AFM cantilever, and immersed within a surfactant solution in water. The sphere was oscillated in the direction normal to the air-water interface, thus generating a flow and an associated hydrodynamic pressure field that could deform the interface. The resulting force exerted on the sphere was measured as a function of the sphere-bubble distance, and found to depend on the bubble size. We also developed a model, coupling axisymmetric lubrication flow and capillary deformations, and accounting for finite-size effects through an asymptotic-matching method. The experimental results were found to be in good agreement with the model, the air-water surface tension being the single free parameter. Finally, from a comparison with independent tensiometry measurements using the Wilhelmy-plate method, we discussed the capacity of our novel method to measure surface tensions robustly. The volume of the liquid required in our method can be as small as tens of microliters. All together, this work paves the way to contactless capillary rheology, with fundamental perspectives in confined soft matter, and practical applications towards micro-monitoring of water contamination. 

\section{Acknowledgements}
The authors thank Elisabeth Charlaix for preliminary discussions, as well as Samir Almohamad for technical assistance on the Wilhelmy-plate calibration experiments. Z. Z. acknowledges financial support from the China Scholarship Council.  Z. Z. and A. M. acknowledge financial support from Agence Nationale de la Recherche (ANR-19-CE30-0012).
\bibliography{biblio}
\end{document}